\pdfoutput=1

\newcommand{\abstracttext}{This work investigates the link between residual entropy and viscosity based on wide-ranging, highly accurate experimental and simulation data.  This link was originally postulated by Rosenfeld in 1977, and it is shown that this scaling results in an approximately monovariate relationship between residual entropy and reduced viscosity for a wide range of molecular fluids (argon, methane, CO$_2$, SF$_6$, refrigerant R-134a (1,1,1,2-tetrafluoroethane), refrigerant R-125 (pentafluoroethane), methanol, and water), and a range of model potentials (hard sphere, inverse power, Lennard-Jones, and Weeks-Chandler-Andersen).  While the proposed ``universal" correlation of Rosenfeld is shown to be far from universal, when used with the appropriate density scaling for molecular fluids, the viscosity of non-associating molecular fluids can be mapped onto the model potentials.  This mapping results in a length scale that is proportional to the cube root of experimentally measureable liquid volume values.}

\newcommand{\acktext}{Arno Laesecke provided ideas beyond Rosenfeld's entropy scaling, contributed his metrological expertise to select the eight molecular fluids, and compiled the most accurate experimental viscosity data for them. He also provided state-of-the-art information about the properties of the hard-sphere potential, the soft-sphere potentials, and suggested to include the viscosity of the Lennard-Jones potential with advice about the most accurate simulation results. The authors also thank: Oliver L\"otgering-Lin, Madlen Hopp and Joachim Gross (University of Stuttgart) for many stimulating discussions and for introduction to this field of research; Chris Muzny (NIST) for project management and for help interfacing with TDE; Allan Harvey (NIST) for discussions of theory and tireless proofreading; Richard Messerly (NIST) for help making sense of soft-sphere potentials; William Krekelberg (NIST) for discussion of molecular simulation and providing simulation results; Marcia Huber (NIST) for providing the experimental data collection and bibliographic information for water; Evan Abramson (University of Washington) for assistance with the conversion of the EOS of Giordano to residual Helmholtz energy;  Monika Thol (Ruhr-Universit\"at Bochum) for help with the Lennard-Jones potential; J. Richard Elliott (University of Akron) for discussions and assistance with model potentials; Yury Fomin (Russian Academy of Sciences) for providing the tabular values of the simulated soft-sphere viscosities; and Eugene Paulechka (NIST) for identification of fluids likely to form hydrogen bonds.}

\documentclass[10pt,twocolumn]{article}

\usepackage[utf8]{inputenc}

\newcommand{\dropcap}[1]{#1}
\newcommand{\acknow}[1]{}
\newcommand{\showacknow}[1]{}

\usepackage{graphicx}
\usepackage{amsmath}
\usepackage{amsfonts}
\usepackage{amssymb}
\usepackage{color}
\usepackage{lscape}
\usepackage{float}
\usepackage{textcomp}
\usepackage{gensymb}
\usepackage{graphicx}
\usepackage{booktabs}
\usepackage{float}
\usepackage{longtable}
\usepackage{listings}
\usepackage[hmargin=0.5in,vmargin=1in]{geometry}

\usepackage{zref-xr,zref-user}
\zxrsetup{verbose}
\zexternaldocument*{SI}
\usepackage[capitalise]{cleveref}

\usepackage{subcaption}
\author{Ian H. Bell}

\title{Probing the link between residual entropy and viscosity of molecular fluids and model potentials}

\begin{document}
\maketitle
\section*{Abstract}
\abstracttext

\newcommand{\Tc}{\ensuremath{T_{\rm{c}}}}
\newcommand{\kB}{\ensuremath{k_{\rm B}}}
\newcommand{\NA}{\ensuremath{N_{\rm A}}}
\newcommand{\rhoND}{\ensuremath{\rho_{\rm ND}}}
\newcommand{\sr}{\ensuremath{s^{\rm{r}}}}
\newcommand{\alphar}{\ensuremath{\alpha^{\rm{r}}}}
\newcommand{\etar}{\ensuremath{\eta^{\rm{r}}}}
\newcommand{\deriv}[3]{\ensuremath{  \left(\dfrac{\partial #1}{\partial #2}\right)_{#3} }}

\dropcap{I}n 1977 Rosenfeld \cite{Rosenfeld-PRA-1977} postulated a quasi-universal relationship between reduced transport properties and the reduced residual entropy.  This analysis was based on the analysis of simulation data for hard spheres, the one-component plasma, and the Lennard-Jones 12-6 model potential in the \textit{liquid phase only}.  This scaling, here referred to as the \textit{Rosenfeld scaling}, was of the form
\begin{equation}
\label{eq:Rosenfeldscaling}
\frac{\eta}{\eta^{\rm R}} = f\left(-\frac{s^{\rm r}}{R}\right),
\end{equation}
where the reducing viscosity $\eta^{\rm R}$, in the same units as $\eta$, is given by
\begin{equation}
\label{eq:etaR}
\eta^{\rm R} = \rho^{2/3}_{\rm N}\sqrt{m\kB T},
\end{equation}
which is obtained by scaling the viscosity in units of Pa$\cdot$s (with dimensions of mass/(length$\times$time)) by the appropriate dimensional scaling parameters (for Newtonian dynamics, mass: $m$, time: $\rho_{\rm N}^{-1/3}\sqrt{m/(\kB T)}$, length: $\rho_{\rm N}^{-1/3}$ )\cite{Gnan-JCP-2009-PartIV,Dyre-JPCM-2016}.  The parameter $\rho_{\rm N}$ is the number density, not to be confused with the molar density $\rho$, $m$ is the mass of one particle or molecule in kg, \kB\ is the Boltzmann constant in J mol$^{-1}$, $T$ is the temperature in kelvins, and $-s^{\rm r}/R$ is the reduced residual entropy.  For more on the selected unit system and nomenclature, see the supporting information (SI) in \zref{sec:units}.

Twenty-two years later, in 1999, Rosenfeld \cite{Rosenfeld-JPCM-1999} proposed the ``universal" correlation for viscosity given by
\begin{equation}
\label{eq:Rosenfeldinvars}
\frac{\eta}{\eta^{\rm R}} = 0.2\exp\left(-0.8\frac{s^{\rm r}}{R}\right).
\end{equation}

Over the last few years, a theoretical basis for the scaling effects that Rosenfeld saw four decades ago has been developed with isomorph theory\cite{Bailey-JCP-2008-PartI,Bailey-JCP-2008-PartII,Schroder-JCP-2009-PartIII,Gnan-JCP-2009-PartIV,Schroder-JCP-2011-PartV,Dyre-JPCM-2016}.  This theory stipulates that the viscosity scaled in the manner of Eq. (\ref{eq:Rosenfeldscaling}) should be invariant along lines of constant residual entropy if there is a high degree of correlation between fluctuations in the virial of the system and fluctuations in its intermolecular potential energy.  A fluid that follows this behavior, even in \textit{some} of its phase space, is referred to as an R-simple (Roskilde simple) fluid \cite{Dyre-JPCM-2016}.  No molecular fluids are truly perfectly correlating in the R-simple sense, and furthermore, this R-simple scaling may only apply in part of the liquid domain, but this is a powerful theoretical tool to understand the dynamic behavior of molecular fluids.  

Rosenfeld scaling can also be applied to dynamic properties like diffusivity; there are now a number of studies focused on Rosenfeld scaling of diffusivity from molecular simulation (see for instance \cite{Dzugutov-NATURE-1996,Krekelberg-PRE-2009-Anomalous,Krekelberg-PRE-2009-Generalized,Agarwal-JPCB-2007,Chopra-JPCB-2010,Goel-JCP-2008}) due to the ease with which self-diffusion can be extracted from the results of molecular simulations.  Studies considering the entropy scaling of experimental diffusivity measurements are growing in number as well \cite{Gerek-IECR-2010,Chopra-JPCB-2010, Liu-CES-1998,Hopp-IECR-2018-selfdiffusion}.  As is highlighted by \cite{Gerek-IECR-2010,Chopra-JPCB-2010}, and also seen in this work in the case of viscosity, one of the limitations of the Rosenfeld scaling applied to self-diffusion is that unique curves are in general obtained for each species studied.  The residual entropy corresponding states approach proposed in this work should also apply to self-diffusion, allowing for harmonization of the self-diffusion studies that have been carried out thus far.

The Rosenfeld scaling of viscosity has been comparatively less studied. Abramson\cite{Abramson-JPCB-2014, Abramson-HPR-2011-argon, Abramson-PRE-2011-methane, Abramson-PRE-2009, Abramson-PRE-2007, Abramson-JCP-2005} was one of the first to consider the Rosenfeld scaling of his experimental viscosity data at very high pressures.  Since then, modified Rosenfeld scaling of viscosity (reducing by the dilute-gas viscosity rather than \cref{eq:etaR}) has also been successfully investigated \cite{Novak-IJCRE-2011-Viscosity,Novak-IJCRE-2011-ViscosityDiffusion,Novak-IECR-2013a,Novak-IECR-2013b,Novak-IECR-2015,LotgeringLin-IECR-2015,LotgeringLin-IECR-2018}.

This work investigates the hypothesis that the Rosenfeld-scaled viscosity should in general be invariant along lines of residual entropy, as is proposed by isomorph theory.  The most comprehensive study to date of this hypothesis based upon viscosities obtained from experimental measurements of molecular fluids and molecular simulation of model potentials has been carried out.  The nearly monovariate relationship between reduced viscosity and residual entropy in the liquid phase, where simple fluids are approximately R-simple, is shown.  Furthermore, this monovariate scaling is shown to apply surprisingly well to hydrogen-bonding fluids approaching the melting line.  Network forming (hydrogen bonding) tends to destroy the R-simple character of the fluid and result in a non-monovariate scaling between reduced viscosity and residual entropy.

The model potentials show the same monovariate dependency of reduced viscosity on the residual entropy as the molecular fluids, and deviate from this behavior in the same ways.  The scaling of the molecular fluids and the model potentials are collapsed by a residual entropy corresponding states approach.  This residual entropy corresponding states is unlike the classical corresponding states\cite{Pitzer-JCP-1939} in which the thermodynamic states, relative to the respective critical points, are equated.  In this case, the residual entropy (the measure of structure of the fluid phase) is the parameter that must be corresponding for dynamic states to be equivalent.  That is to say: \textit{residual entropy is the scaling parameter that connects the thermodynamic and transport properties of dense molecular fluids}. 

\section{Molecular fluids}
\label{sec:molecularfluids}

The term ``molecular fluid" is used in this work to differentiate from model intermolecular potentials that are useful theoretical models but are not experimentally accessible in a laboratory.  The study of molecular fluids in this section is indebted to the work of the experimental transport property community; without their tireless work, this study would not have been possible.

\subsection{Fluid selection}

Molecular fluids for this study were selected according to the availability of:
\begin{enumerate}
\setlength\itemsep{-0.5em}
\item a significant body of high-quality experimental viscosity data covering most of the liquid, gas, and supercritical states, and
\item a well-constructed equation of state for the thermodynamic properties that yields high-fidelity predictions of the residual entropy over the entire fluid range.
\end{enumerate}  

Unfortunately, there are not many fluids (perhaps 30) that meet these requirements.
The selected molecular fluids represent the following classes:
\begin{itemize}
\setlength\itemsep{-0.5em}
\item monatomic gas (argon),
\item predominantly repulsive molecules (methane, carbon dioxide, and sulfur hexafluoride),
\item halogenated refrigerants (R-134a and R-125) with electrostatic interactions due to polarity,
\item strongly associating fluids (methanol and water)
\end{itemize}  

Table \ref{tab:EOS_ref} lists the equations of state that were employed in this work.  All of the equations of state are multiparameter reference equations.  The NIST REFPROP thermophysical property library \cite{LEMMON-RP10} was used to carry out all the calculations.  The equations of state in REFPROP have been critically assessed and deemed to be the most reliable for the given fluid and all have been published in the literature.

\begin{table}[H]
\caption{Equations of state used in this work \label{tab:EOS_ref}}
\centering
\begin{tabular}{ccccc}
\hline\hline
Common name & EOS & $T_{\rm max}$ (K) & $p_{\rm max}$ (MPa) \\
\hline
Argon & \cite{Tegeler-JPCRD-1999} & 2000 & 1000 \\
Methane & \cite{Setzmann-JPCRD-1991} & 625 & 1000 \\
SF$_6$ & \cite{Guder-JPCRD-2009} & 625 & 150 \\
CO$_2$ & \cite{Span-JPCRD-1996}$^a$ & 2000 & 800 \\
R-134a & \cite{TillnerRoth-JPCRD-1994} & 455 & 70 \\
R-125 & \cite{Lemmon-JPCRD-2005} & 500 & 60 \\
Methanol & \cite{deReuck-BOOK-1993} & 620 & 800 \\
Water& \cite{Wagner-JPCRD-2002} & 2000 & 1000 \\
\hline\hline
\end{tabular} \\
a: The equation of state of Giordano \textit{et al.} \cite{Giordano-JCP-2006} is used above 800 MPa instead of that of Span and Wagner\cite{Span-JPCRD-1996}. \\
\end{table}

If temperature and pressure are known for the experimental state point, the density is iteratively obtained from the equation of state.  With the exception of carbon dioxide, for which the equation of state of \cite{Giordano-JCP-2006} (see the SI Section \zref{sec:Giordano} for the use of this EOS) was used above the maximum pressure of the equation of state of \cite{Span-JPCRD-1996}, any measurements carried out at pressures above the stated maximum pressure of the EOS were excluded to avoid errors associated with extrapolation. 

In the SI, Fig. \zref{fig:coverage_overview} shows the coverage of the experimental viscosity data available for the studied fluids, and the limits of the equations of state for these fluids.  This figure demonstrates that there is significant disparity in data coverage, even among the best-studied fluids. 

\subsection{Evaluation of $\sr$}
\label{sec:evaluationsr}

The state-of-the-art equations of state for molecular fluids are Helmholtz-energy-explicit with temperature and density as independent variables.  In these formulations, the molar Helmholtz energy $a$ is expressed as a sum of the ideal-gas $a^0=RT\alpha^0$ and residual $a^{\rm r}=RT\alpha^{\rm r}$ contributions, given as
\begin{equation}
\alpha(\tau,\delta) = \frac{a}{RT} = \alpha^0(\tau,\delta)+\alphar(\tau,\delta),
\end{equation}
where the independent variables are the reciprocal reduced temperature $\tau=\Tc/T$ and the reduced density $\delta=\rho/\rho_{\rm c}$, and $\Tc$ and $\rho_{\rm c}$ are the critical temperature and molar density respectively.

Expressed in terms of derivatives of $\alpha$, the molar entropy $s=-(\partial a/\partial T)_{\rho}$ is given by
\begin{equation}
\label{eq:s}
\frac{s}{R}= \tau \left[\left(\frac{\partial \alpha^0}{\partial \tau}\right)_{\delta}+ \left(\frac{\partial \alphar}{\partial \tau}\right)_{\delta} \right]-\alpha^0-\alphar.
\end{equation}
and the residual entropy $\sr$ is the part of \cref{eq:s} that is based only on $\alphar$ and its derivatives, resulting in
\begin{equation}
\label{eq:srRmolecular}
\frac{\sr}{R} = \tau\left(\frac{\partial \alphar}{\partial \tau}\right)_{\delta} - \alphar.
\end{equation}

For additional details of the the use of multiparameter EOS, the reader is directed to the literature \cite{Span-BOOK-2000,Bell-IECR-2014,Kunz-BOOK-2007}.

The residual entropy should not be confused with the term ``excess" entropy \cite{Marsh-JCED-1997,Gerek-IECR-2010}, which refers to differences of mixture thermodynamics from ideal solution behavior.

\begin{figure}[htb]
\centering
\includegraphics[width=3.3in]{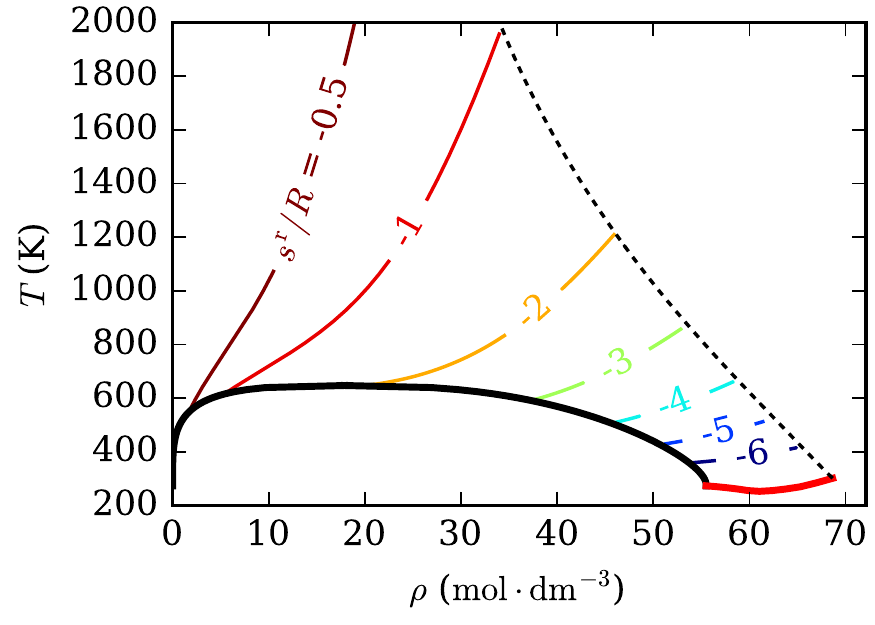}
\caption{Contours of the residual entropy $\sr/R$ for water from the equation of state of Wagner and Pru\ss \cite{Wagner-JPCRD-2002}.  The dashed curve is the line of maximum pressure of the equation of state, the solid red curve is the melting curve, and the solid black curve is the vapor-liquid co-existence curve (the binodal). \label{fig:WATER_contours}}
\end{figure}

The residual entropy is defined as the part of the entropy that arises from the interactions among particles or molecules.  This contribution is negative due to repulsive and attractive interactions that increase the structure beyond that of the non-interacting ideal gas\cite{Ingebrigtsen-JPCB-2012}. To illustrate this property, \cref{fig:WATER_contours} shows contours of the reduced residual entropy for ordinary water, where $-\sr/R$ is evaluated from the equation of state of Wagner and Pru\ss \cite{Wagner-JPCRD-2002}.  In the zero-density limit, $-\sr/R$ is zero (no increase in structure caused by molecular interactions), and as the density increases, so does $-\sr/R$.  The maximum value for $-\sr/R$ is found along the melting line at the maximum pressure of the equation of state; this can be intuitively understood as the state within the fluid domain where the fluid is most structured.  

\subsection{Data Analysis}

\begin{figure*}[ht]
\centering
\includegraphics[width=6.5in]{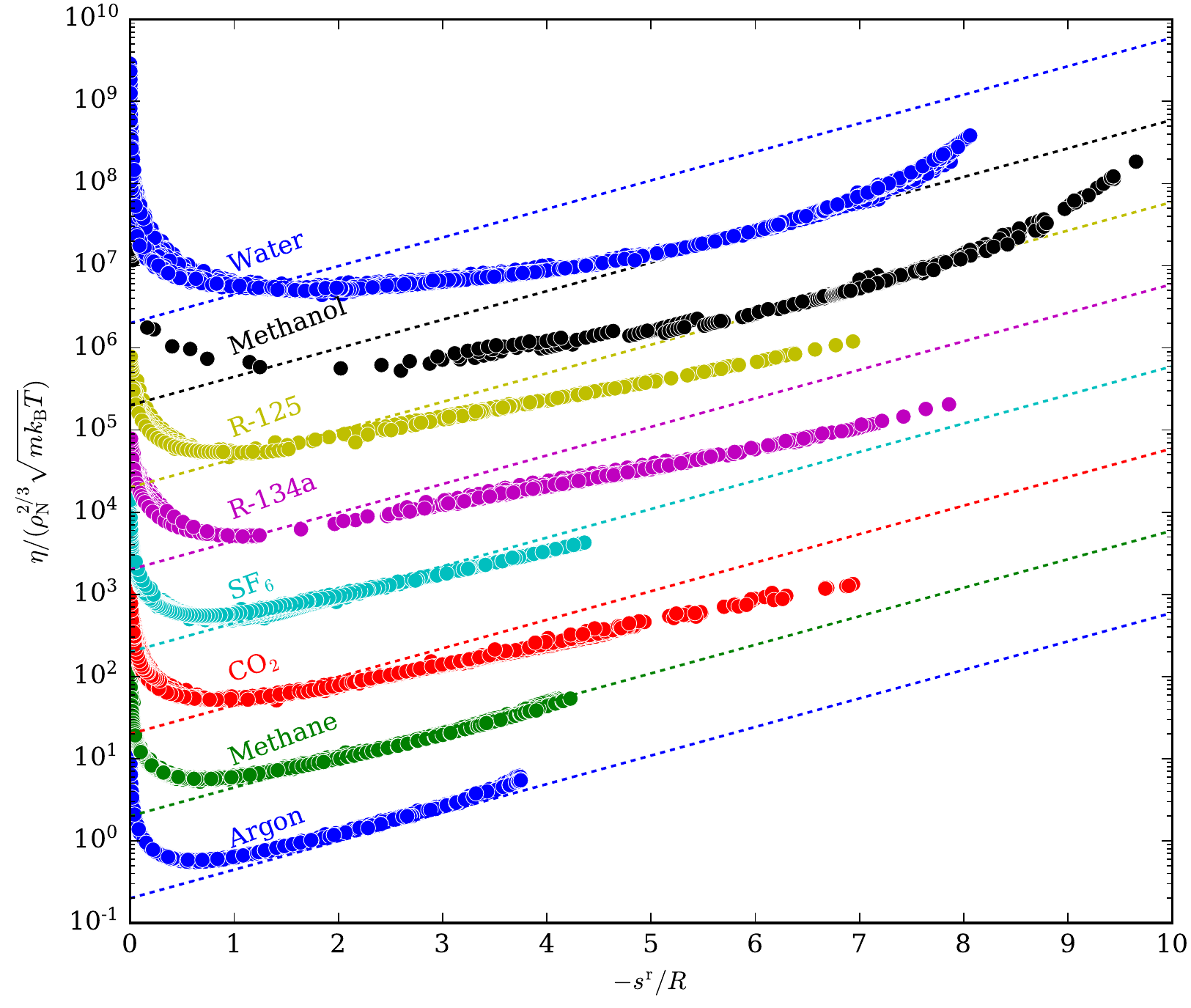}
\caption{Overview of relationship between reduced viscosity and residual entropy for the molecular fluids from a total of 12987 experimental data points.  The dashed line represents the ``universal" scaling law of Rosenfeld \cite{Rosenfeld-JPCM-1999}.  The data are vertically stacked by multiplying by increasing powers of 10.   \label{fig:sr_overview_no_rho_etaRosenfeld}}
\end{figure*}

Experimental viscosity data were curated for a selection of fluids that experience more complex interactions than the simple model fluids investigated by Rosenfeld\cite{Rosenfeld-PRA-1977}.  For each experimental data point, the molar density was determined, either taken directly from the measurement or from an iterative thermodynamic calculation of the equation of state.  The residual entropy was then evaluated at the specified molar density and temperature as described in \cref{sec:evaluationsr}.

Figure \ref{fig:sr_overview_no_rho_etaRosenfeld} shows the experimental viscosity data for the eight molecular fluids under study, with the Rosenfeld ``universal" relationship overlaid for each fluid. The viscosity is reduced in the same manner as proposed by Rosenfeld \cite{Rosenfeld-PRA-1977}.  The data for each of the fluids in these scaled coordinates has a characteristic, and roughly similar, shape.  The data for other molecular fluids (investigated but not discussed in this paper) also have the same shape.  

For the fluids that are Lennard-Jones-like (e.g., argon or methane), the ``universal" correlation of Rosenfeld captures the correct qualitative relationship between the viscosity and the residual entropy at liquid-like conditions ($-\sr/R \gtrsim  1$) at densities that are not ``too high". As the intermolecular interactions qualitatively increase in intensity, (i.e., for the associating fluids), the Rosenfeld ``universal" relationship does not agree either qualitatively or quantitatively in the liquid-like phase.  

Costigliola et al. \cite{Costigliola-JCP-2018} and others \cite{Ingebrigtsen-JPCB-2012,Albrechtsen-PRB-2014,Hummel-PRB-2015} suggest that water (and other associating fluids) should not have a monovariate viscosity scaling in terms of residual entropy in the liquid phase due to the presence of hydrogen-bonding networks.  \cref{fig:sr_overview_no_rho_etaRosenfeld} shows that water does in fact demonstrate an approximate collapse of the reduced viscosity surface into a monovariate dependency on the reduced residual entropy, with the exception of states approaching the melting line, where the analysis of Ruppeiner and co-authors (see \cref{sec:liquids}) suggests a means of identifying the presence of hydrogen-bonding networks from a high accuracy equation of state.  

\subsubsection{Liquids}
\label{sec:liquids}

For liquid-like states ($-\sr/R \gtrsim  1$), the experimental data for each non-associating molecular fluid (aside from some scatter in the experimental measurements) collapse onto master curves -- a monovariate functional dependence.  The curvature in semi-log coordinates differs depending on the intermolecular interactions.  In the case of argon, methane, SF$_6$, CO$_2$, and the refrigerants R-134a and R-125, the liquid-like scaling is roughly linear in semi-log coordinates.  The experimental data for these fluids (particularly for SF$_6$ and CO$_2$ and less so for R-134a and R-125) extend to the melting line (see Fig. \zref{fig:coverage_overview} in the SI). 

For the associating fluids methanol and water, a more complicated functional dependence is seen, particularly at large values of $-\sr/R$.  While the reduced viscosity data are still a nearly monovariate function of the residual entropy, the curvature of the data increases at higher values of $-\sr/R$.  The pronounced increase in curvature can be ascribed to the presence of transient structures in the fluid caused by hydrogen-bonding networks in the bulk liquid phase.  This pronounced curvature in the Rosenfeld-scaled viscosity is consistent with the behavior identified by other authors for diffusivity\cite{Chopra-JPCB-2010} and viscosity\cite{Abramson-PRE-2007}  of water.  The thermodynamic states where these networks are present can be identified by states with positive Riemannian curvature \cite{Ruppeiner-RMP-1995,Ruppeiner-PLA-2015,Branka-PRE-2018,Mausbach-PRE-2018}.

\begin{figure}[htb]
\centering
\includegraphics[width=3.3in]{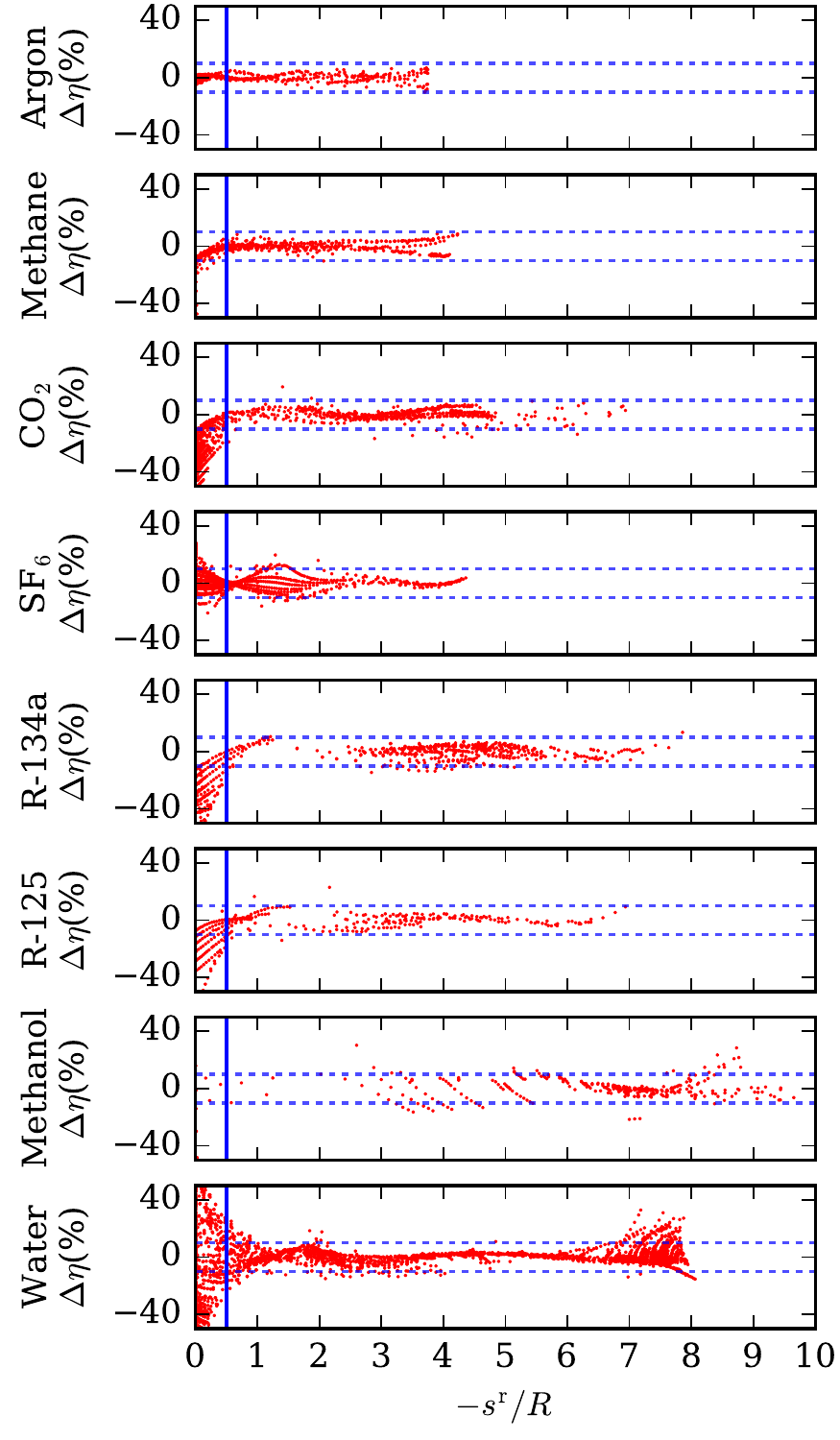}
\caption{Deviations from monovariability between data and fits from \cref{eq:monofit}. Deviation term is given by $\Delta\eta=100\times(\eta_{\rm fit}/\eta_{\rm exp}-1)$ and dashed lines represent $\pm 10$\%.  The correlation is fit for $-s^{\rm r}/R > 0.5$, values below the range of the fit are impacted by the extrapolation behavior of the fit and non-monovariate scaling. \label{fig:srfit_deviations}}
\end{figure}

In order to assess the monovariability of the relationship between the reduced viscosity and the residual entropy, polynomial correlations for each fluid from $-s^{\rm r}/R =0.5$ up to the melting curve of the fluid were developed in this work.  The correlations were of the form
\begin{equation}
\label{eq:monofit}
\ln\left[\frac{\eta}{\rho_{\rm N}^{2/3}\sqrt{m\kB T}} \cdot \left(-\frac{s^{\rm r}}{R}\right)^{2/3}\right] = \sum_{i} c_i\left(-\frac{s^{\rm r}}{R}\right)^i.
\end{equation}
Multiplication of the reduced viscosity by $(-s^{\rm r}/R)^{2/3}$ was used to remove the divergence in the dilute-gas limit.  The coefficients of the polynomial fits are in the SI (Table \zref{tab:srfit_fits}).  \Cref{fig:srfit_deviations} shows the deviations between the experimental data points and the fits obtained from \cref{eq:monofit}.  Even though the absolute deviations are as much as 35 \% for the hydrogen-bonding fluids in the compressed liquid due to the breakdown in monovariability caused by hydrogen bonding, the average absolute deviation (AAD) for each fluid is less than 4.3\% for ($-\sr/R > 0.5$).  This shows that the relationship between reduced viscosity and residual entropy is indeed approximately monovariate (except for the associating fluids).

These liquid-like data do not directly refute the analysis of Rosenfeld, who merely proposed a monovariate relationship between reduced viscosity and residual entropy. The results in \cref{fig:sr_overview_no_rho_etaRosenfeld} and \cref{fig:srfit_deviations} indicate that this monovariate relationship is present, even if the relationship might be different for each fluid.  As shown in \cref{sec:squarecircle}, the mapping onto the results of model intermolecular potentials can reduce the data so that non-associating fluids can all be collapsed onto a single curve with one adjustable parameter.  

\subsubsection{Gas}

In the gaseous domain where $-\sr/R \lesssim  1$, there is a pronounced deviation from monovariate scaling, as is visible in \cref{fig:sr_overview_no_rho_etaRosenfeld}, and more readily seen in the detailed view of this region in the SI (Fig. \zref{fig:scaled_data_12987pts_gas_zoomed}).  This figure demonstrates two deficiencies in the scaling proposed by Rosenfeld:
\begin{itemize}
\setlength\itemsep{-0.3em}
\item The scaling diverges at zero density (where $-\sr/R =0$)
\item The gaseous region does not reduce to a monovariate dependence of reduced viscosity with $-\sr/R$
\end{itemize}

Other authors \cite{Novak-IJCRE-2011-Viscosity,Novak-IJCRE-2011-ViscosityDiffusion,Novak-IECR-2013a,Novak-IECR-2013b,Novak-IECR-2015,Bell-PURDUE-2016-eta,LotgeringLin-IECR-2015,LotgeringLin-IECR-2018} have proposed alternative residual-entropy-based schemes that are more successful at scaling the viscosity in the dilute gas limit, but they introduce significant deviations from monovariate scaling in the compressed liquid phase for small non-associating molecules and in the gaseous phase for associating molecules. For that reason the alternative entropy scalings are not discussed in this work; further work will investigate potential means of reconciling these different scaling approaches.  

The isomorph theory describes \textit{why} the reduced viscosity should not be a monovariate function of the residual entropy in the gas phase.  In the gaseous region, the motion of the molecules is predominantly ballistic, aside from the infrequent interactions between molecules via collision.  Therefore, the fluid \textit{should not be} R-simple, isomorph scaling should be invalid, and reduced viscosity should not be a monovariate function of residual entropy. 

\section{Model potentials}
\label{sec:modelpotentials}

Model intermolecular potentials, and the simulation results that are obtained from these potentials, have much to teach us about transport properties of molecular fluids.  After some general information, this section covers four model potentials.

The viscosity of single-site models in molecular simulations is in general given in the form $\eta^*=\eta\sigma^2/\sqrt{m\varepsilon}$ \cite{Meier-JCP-2004}, in terms of the reduced temperature $T^*=T/(\varepsilon/\kB)$ and reduced density $\rho^*=\rhoND\sigma^3$, where $\varepsilon$ is an energy scale, $\sigma$ is a length scale, and $\rhoND$ is the number density in m$^{-3}$; for more information on working in molecular simulation units, see \cite{Lustig-MP-2012}.  

\subsection{Hard sphere}

The hard sphere model potential is a particularly simple one; rigid spherical particles have ballistic trajectories until they collide with another particle.  The reduced viscosity of the hard-sphere potential, as well as its associated residual entropy, can each be obtained as a function of the packing fraction $\zeta = \pi \rho^*/6$.  The parameter $\zeta$ is \textit{not} a function of temperature, but only a function of density. In the SI (Fig. \zref{fig:hard_sphere}) is a graphical representation of the scaling for the hard sphere, and the curve of the reduced viscosity versus the residual entropy is also shown in \cref{fig:overlay_simulations}.  The shape of the viscosity versus $-s^{\rm r}$ curve in scaled coordinates bears significant but imperfect resemblance to that of the argon data in \cref{fig:sr_overview_no_rho_etaRosenfeld}. 
 
\subsection{Inverse-power pair potential}

Real molecules are not rigid; they are more rubber ball than billiard ball.  As a result, it is more reasonable to treat molecules as soft spheres than hard spheres.  The inverse-power pair potential (IPP) is a repulsive potential commonly used to model fluids with soft repulsive interactions given by $U = \varepsilon\left(\sigma/r\right)^{n}$. The density and temperature are not independent for the IPP potential \cite{Tan-MP-2011,Barlow-JCP-2012,Bailey-JCP-2008-PartII}; they are linked via the scaling variable $\gamma = \rho_{\rm N}\sigma^3(T^*)^{-3/n}$.

The ratio of viscosity $\eta$ to $\rho_{\rm N}^{2/3}\sqrt{m\kB T}$  is then (see the SI)
\begin{equation}
\frac{\eta}{\rho_{\rm N}^{2/3}\sqrt{m\kB T}} = \dfrac{\eta^*/(T^*)^{n'}}{\gamma^{2/3}},
\end{equation}
in which $n'=(2/n)+(1/2)$. The simulation for the IPP potential is carried out at specified pairs of $\rho^* = \rho_{\rm N}\sigma^3$ and $T^*=T\kB/\varepsilon$, for which the simulation data are expressed in terms of $\eta^*/(T^*)^{n'}$ as a function of the scaling variable $\gamma$ (see \cite{Fomin-JETPLetters-2012} and Section \zref{sec:IPPdense}).  For the $n=12$ IPP potential, the residual entropy is obtained from integration of the convergent virial expansion given by \cite{Tan-MP-2011}. For other values of $n$, the asymptotically convergent approximation of \cite{Barlow-JCP-2012} is used (see the SI Section \zref{sec:IPPsr} for further description of this method).

\subsection{Lennard-Jones}
\label{sec:LennardJonespotential}

Real fluids interact by both attraction and repulsion (as well as long-range electrostatic interactions); the potential should capture this.  The canonical example of a fluid with both attraction and repulsion is the Lennard-Jones 12-6 potential; it is given by
\begin{equation}
U = 4\varepsilon\left[\left(\frac{\sigma}{r}\right)^{12}-\left(\frac{\sigma}{r}\right)^6\right].
\end{equation}

A number of researchers have carried out molecular simulation on the Lennard-Jones 12-6 potential, and through application of the Green-Kubo formalism, evaluated viscosities \cite{Maginn-LJCMS-2018}.  The coverage of the simulation results for the Lennard-Jones fluid is shown in the SI Fig. \zref{fig:coverage_LJ}.  The most accurate equation of state for the Lennard-Jones 12-6 potential is the one recently developed by Thol \textit{et al.} \cite{Thol-JPCRD-2016}, which is valid up to $T^*=9$ and $p^*=65$, where  $p^*=p\sigma^3/\varepsilon$.  Due to the availability of the multiparameter EOS for the Lennard-Jones 12-6 potential\cite{Thol-JPCRD-2016}, the same methodology for the Lennard-Jones 12-6 potential is applied as with the molecular fluids in \cref{sec:evaluationsr} -- for a given set of $T^*$, $\rho^*$, $\eta^*$ from one simulation, the residual entropy is evaluated from the equation of state as described in \cref{eq:srRmolecular}.

\subsection{WCA}
\label{sec:WCApotential}

Weeks, Chandler, and Andersen (WCA) \cite{Weeks-JCP-1971} proposed a means of deconstructing potentials into reference and attractive contributions.  The reference part of the WCA deconstruction of the Lennard-Jones 12-6 potential results in a fully repulsive potential that has dynamic behavior similar to that of the Lennard-Jones 12-6 potential with shorter-ranged interactions. This reference potential is obtained by truncating the Lennard-Jones 12-6 potential at the location of its minimum value at $r=2^{1/6}\sigma$ and shifting the curve upwards by $\varepsilon$, or
\begin{equation}
U = \left\lbrace 
\begin{array}{cc}
4\varepsilon\left[\left(\dfrac{\sigma}{r}\right)^{12}-\left(\dfrac{\sigma}{r}\right)^6\right] + \varepsilon, & r \leq 2^{1/6}\sigma \\
0, & r > 2^{1/6}\sigma.
\end{array}
\right.
\end{equation}

We refer to the reference part of the WCA deconstruction of the Lennard-Jones 12-6 potential as the "repulsive WCA potential" for concision.

The repulsive WCA potential retains the same intermolecular force between molecules as the Lennard-Jones 12-6 potential (the force between particles is the negative of the derivative of the potential with respect to position) within $r \leq 2^{1/6}\sigma$.  The transport properties of the repulsive WCA potential are similar to those of the Lennard-Jones 12-6 potential, while having thermodynamic properties that are more straightforward to evaluate because the equation of state reduces to a quasi-monovariate function of the effective packing fraction, without a liquid phase or a critical point \cite{Ahmed-PRE-2009}.

\newcommand{\sigmae}{\ensuremath{\sigma_{\rm e}}}
Analogously to the soft-sphere potential, an effective packing fraction (with implicit temperature dependence) is defined by Heyes and Okumara \cite{Heyes-JCP-2006-WCA} by $\zeta_{\rm e} = \pi\rho^*\left(\sigmae/\sigma\right)^3/6$, with the effective particle diameter given by $\sigmae/\sigma = [2/(1+\sqrt{T^*})]^{1/6}$.
Alternative effective particle diameter models are described in the literature \cite{Elliot-FPE-1986,Ghobadi-JCP-2013-PartI,Kolafa-FPE-1994,Verlet-PRA-1972}.  The residual entropy of the repulsive WCA potential is obtained by integration of the empirical compressibility factor model proposed by Heyes and Okumara \cite{Heyes-JCP-2006-WCA} (see description in the SI, Section \zref{sec:WCA}).  There is currently a  scarcity of high-accuracy tabulated viscosity data for the repulsive WCA potential; however, sufficient data exist to develop the empirical correlation given in the SI. Of particular interest are the new simulation results from Krekelberg provided in the SI with permission, Section \zref{sec:BillWCA}..

\subsection{Overview}

\begin{figure}[htb]
\includegraphics[width=3.2in]{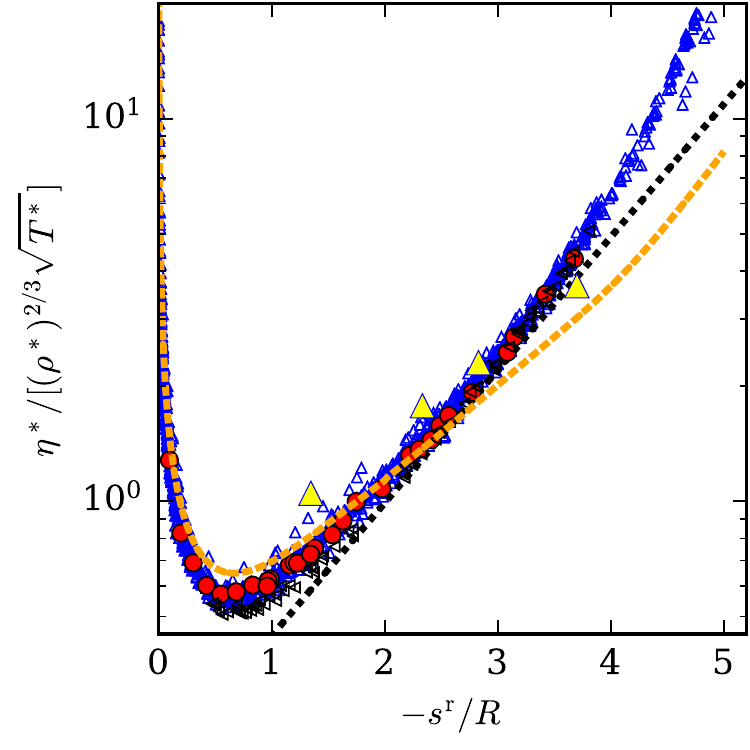}
\caption{Overlaid data for each of the model potentials studied in this work (blue $\triangle$: Lennard-Jones 12-6 potential\cite{Baidakov-JCP-2012,Galliero-IECR-2005,Heyes-PRB-1988,Meier-JCP-2004,Michels-PHYSICAA-1985,Oderji-PRE-2011,Vasquez-IJT-2004}; black $\lhd$: IPP with $n=12$ \cite{Fomin-JETPLetters-2012}; red $\bigcirc$: repulsive WCA potential; yellow $\triangle$: Lennard-Jones data from Ashurst and Hoover\cite{Ashurst-PRA-1975} considered by Rosenfeld \cite{Rosenfeld-PRA-1977}; orange dashed curve: hard sphere (Enskog theory plus correction of \cite{Sigurgeirsson-MP-2003}); black dashed line: correlation from Rosenfeld\cite{Rosenfeld-JPCM-1999}).  A larger version of this figure is available in the SI (Fig. \zref{fig:overlay_simulations_big}). \label{fig:overlay_simulations}}
\end{figure}

\Cref{fig:overlay_simulations} presents the simulation results for all the model potentials included in our study.  These data comprise the corpus of data for the Lennard-Jones 12-6 potential, simulation results for the IPP with $n=12$, results for the repulsive WCA potential, and the curve for the hard-sphere potential.  

The ``universal" scaling of Rosenfeld does not reproduce all of the Lennard-Jones simulation data in the liquid phase.  In the work of Rosenfeld \cite{Rosenfeld-JPCM-1999}, he described good agreement with simulation results for the ``universal" correlation.  In reality, the correlation was compared with a single data set comprising four data points at zero shear rate from Ashurst and Hoover \cite[Table VI]{Ashurst-PRA-1975}; the present data coverage of results on the Lennard-Jones fluid here is far more comprehensive. Rosenfeld's curve of ``universal" scaling might not have been quite right, but with the appropriate caveats, most repulsive-dominated potentials are remarkably consistent in the Rosenfeld scaling framework.

\section{Residual Entropy Corresponding States}
\label{sec:squarecircle}

Figures \ref{fig:sr_overview_no_rho_etaRosenfeld} and \ref{fig:overlay_simulations} demonstrate a remarkable similarity for the non-associating fluids.  The primary difference between fluids and potentials is the scaling of the residual entropy.  Therefore, a means of connecting the molecular fluids and the model potentials is needed.  This link is formed through the use of residual entropy corresponding states.

It is possible to map from experimental units into simulation units $T^*$, $\rho^*$, etc. by adjusting the parameters $\varepsilon/\kB$ and $\sigma$.  Carrying out the appropriate cancellation results in
\begin{equation}
\frac{\eta}{\rho_{\rm N}^{2/3}\sqrt{m\kB T}} = \frac{\eta^*}{\left(\rho^*\right)^{2/3}\sqrt{T^*}}
\end{equation}
and, therefore, scaling properties from number density to $\rho^*$, from temperature to $T^*$, and from viscosity to $\eta^*$ will not change the Rosenfeld-reduced viscosity.  On the other hand, modifying $\varepsilon/\kB$ and $\sigma$ adjust the residual entropy.  

At this point, it is necessary to determine:
\begin{itemize}
\setlength\itemsep{-0.3em}
\item the most appropriate reference potential
\item a set of values for $\sigma$ for the mapping from a molecular fluid to a reference potential
\end{itemize}

While the Lennard-Jones potential is appealing as a model potential, its use as the reference system for molecular fluids is problematic because 
\begin{itemize}
\setlength\itemsep{-0.3em}
\item the Lennard-Jones 12-6 potential behaves like a molecular fluid and has a liquid phase, and also no convenient scaling variable such as the $\gamma$ of the IPP potential,
\item the equation of state for the Lennard-Jones 12-6 potential has areas in the unstable region between the spinodals where non-physical residual entropies are obtained (see the SI),
\item no highly accurate viscosity correlation for the Lennard-Jones 12-6 potential exists, though several empirical viscosity correlations of poorer accuracy are available in the literature \cite{Woodcock-AICHEJ-2006, Rowley-IJT-1997}.
\end{itemize} 
For these reasons, scaling onto the repulsive WCA potential was chosen; the repulsive WCA potential has a compressibility factor that is a monovariate function of the thermodynamic scaling parameter $\zeta_{\rm e}$, and the simulation results for the viscosity of the repulsive WCA potential lie within the range of results from the Lennard-Jones simulations (see \cref{fig:overlay_simulations}). Mapping the properties onto the $n=12$ IPP potential was slightly less successful, as described in the SI (Fig. \zref{fig:simscaled_non_hydrogen}).  The mapping onto the hard-sphere potential was also carried out with the same methodology.  The hard-sphere mapping was not successful, as is shown in the SI (Fig. \zref{fig:simscaled_non_hydrogen_HS}).

The value of $\varepsilon/\kB$ was set equal to the critical temperature of the molecular fluid divided by 1.32 ($T_{\rm c}^*=1.32$ for the Lennard-Jones equation of state\cite{Thol-JPCRD-2016}; the repulsive WCA potential is fully repulsive and therefore does not have a critical point) and  $\sigma$ was left as an adjustable parameter.  In this way, corresponding states between the Lennard-Jones analog (the repulsive WCA potential) and the molecular fluid are enforced.  Values of $\varepsilon/\kB=\Tc$ and $\varepsilon/\kB=\Tc/0.7$ were also considered, as described in the SI; there is a very weak dependence of $\sigma$ on $\varepsilon/\kB$.

\begin{table}
\caption{Optimized values of $\sigma$ for the eight molecular fluids included in this study.  Units of all variables are $10^{-10}$ m (\AA)\label{tab:sigma_values}}
\centering
\begin{tabular}{cccccc}
\toprule
    fluid &  $v_{\rm N, triple}^{1/3}$ & $v_{\rm N, 0.8\Tc}^{1/3}$ &  $v_{\rm N, crit}^{1/3}$ &  $\sigma_{\rm IPP}$ &  $\sigma_{\rm WCA}$ \\
\midrule
Water & 3.104 & 3.334 & 4.529 & 3.084 & 2.973\\
Argon & 3.615 & 3.855 & 4.985 & 3.676 & 3.476\\
CO$_2$ & 3.966 & 4.081 & 5.387 & 3.943 & 3.724\\
Methane & 3.901 & 4.226 & 5.471 & 3.957 & 3.733\\
Methanol & 3.892 & 4.311 & 5.739 & 4.178 & 3.950\\
R-134a & 4.745 & 5.204 & 6.917 & 5.095 & 4.830\\
SF$_6$ & 5.094 & 5.250 & 6.888 & 5.244 & 4.995\\
R-125 & 4.909 & 5.314 & 7.030 & 5.265 & 4.992\\
\bottomrule
\end{tabular}
\end{table}

\label{sec:softspherepotential}
Each fluid was mapped onto the residual entropy of the reference potential.  In order to do this, a one-dimensional optimization of $\sigma$ was carried out to minimize the difference between the Rosenfeld scalings at liquid-like states.  The approach is as follows:
\begin{enumerate}
\setlength\itemsep{-0.3em}
\item For a given molecular fluid experimental datapoint for which $-s^{\rm r}/R > 1$, calculate the reduced quantity $\eta/[\rho_{\rm N}^{2/3}\sqrt{m\kB T}]$.  
\item At the same value of reduced viscosity for the repulsive WCA correlation, calculate the corresponding value of $\zeta_{\rm e}$ for the repulsive WCA potential; the correlation is monotonic. 
\item From $\zeta_{\rm e}$, calculate $\rho^*$ for the given $\sigmae/\sigma$ from $\rho^*=6\zeta_{\rm e}(\sigmae/\sigma)^3/\pi$, and then obtain $\sigma=(\rho^*/\rho_{\rm N})^{1/3}$.
\end{enumerate}
The median value of $\sigma$ among all the experimental data points for which a value of $\sigma$ is successfully obtained is retained; the median $\sigma$ was used in order to avoid the influence of outliers.  It may not be possible to obtain the value for $\sigma$ if $\eta/[\rho_{\rm N}^{2/3}\sqrt{m\kB T}]$ is below the minimum value of $\eta/[\rho_{\rm N}^{2/3}\sqrt{m\kB T}]\approx 0.57$ that can be achieved for the repulsive WCA potential.  Once the value of $\sigma$ has been determined for a molecular fluid, this value can then be used to scale all the experimental data into the ``simulation" units of $T^*$, $\rho^*$, $\eta^*$. 
\begin{figure}[htb]
\centering
\includegraphics[width=3.3in]{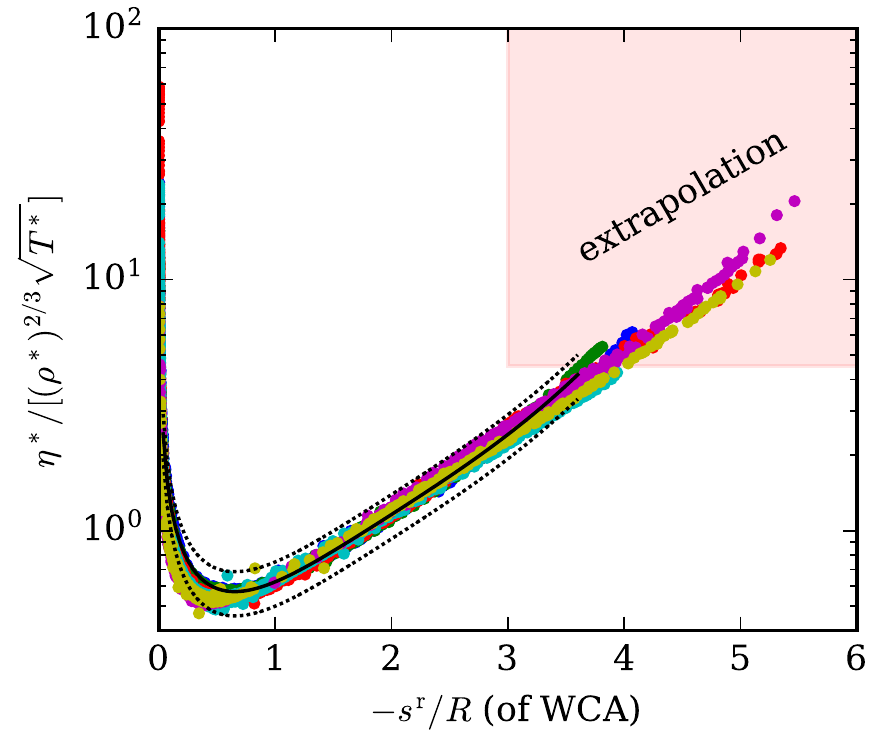}
\caption{Scaled experimental data mapped onto the repulsive WCA potential for the non-associating fluids argon, methane, CO$_2$, SF$_6$, R-134a, and R-125.  The residual entropies are evaluated for the repulsive WCA potential.  The solid line is the correlation for the repulsive WCA potential, and the dashed lines show $\pm20\%$.\label{fig:simscaled_non_hydrogen_WCA}}
\end{figure}

\begin{figure}[htb]
\centering
\includegraphics[width=3.3in]{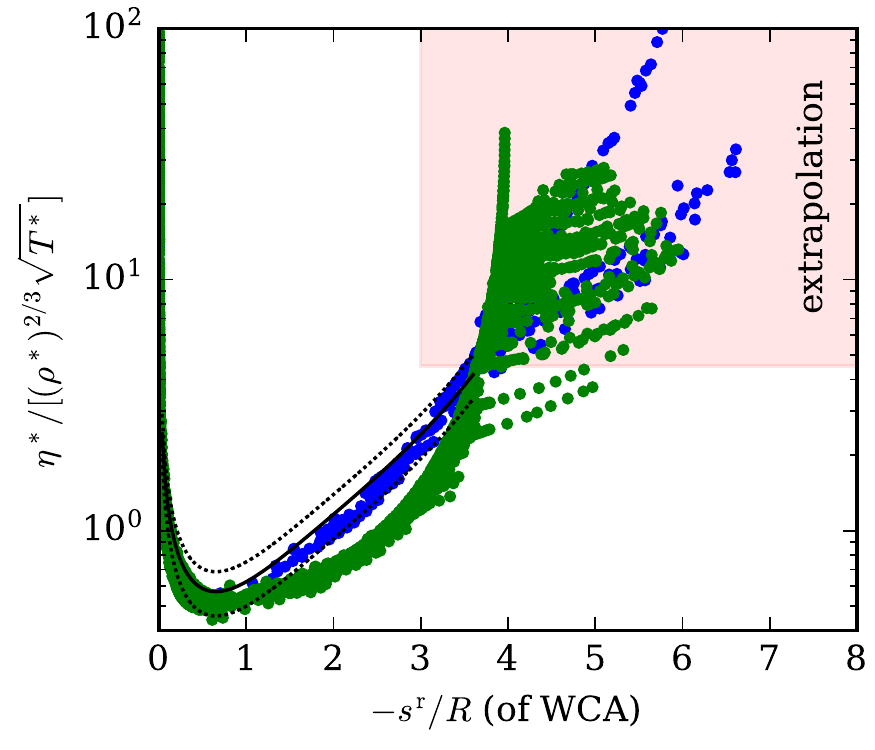}
\caption{Scaled experimental data mapped onto the repulsive WCA potential for the hydrogen-bonding fluids methanol and water. The residual entropies are evaluated for the repulsive WCA potential. The solid line is the correlation for the repulsive WCA potential, and the dashed lines show $\pm20\%$.  The blue markers are for methanol, and the green ones for water. \label{fig:simscaled_hydrogen_WCA}}
\end{figure}

This approach was carried out for the eight molecular fluids discussed in this study, and the obtained values of $\sigma$ are given in \cref{tab:sigma_values}.   \Cref{fig:simscaled_non_hydrogen_WCA} shows the scaled experimental data for the non-associating fluids; the results for assocating fluids are shown in \cref{fig:simscaled_hydrogen_WCA}.  In the case of the non-associating fluids, the qualitative agreement is surprisingly good; with the appropriate scaling, all experimental data can be closely mapped onto a master curve given by the repulsive WCA correlation.  The repulsive WCA model potential does not perfectly match the Rosenfeld-scaled experimental data mapped onto the repulsive WCA potential, and it is evident that although the majority of the data in the liquid phase can be predicted within 20\% (the dashed lines), the curvature of the mapped experimental data does not perfectly match the curvature of the correlation.

\begin{figure}[htb]
\centering
\includegraphics[width=3.3in]{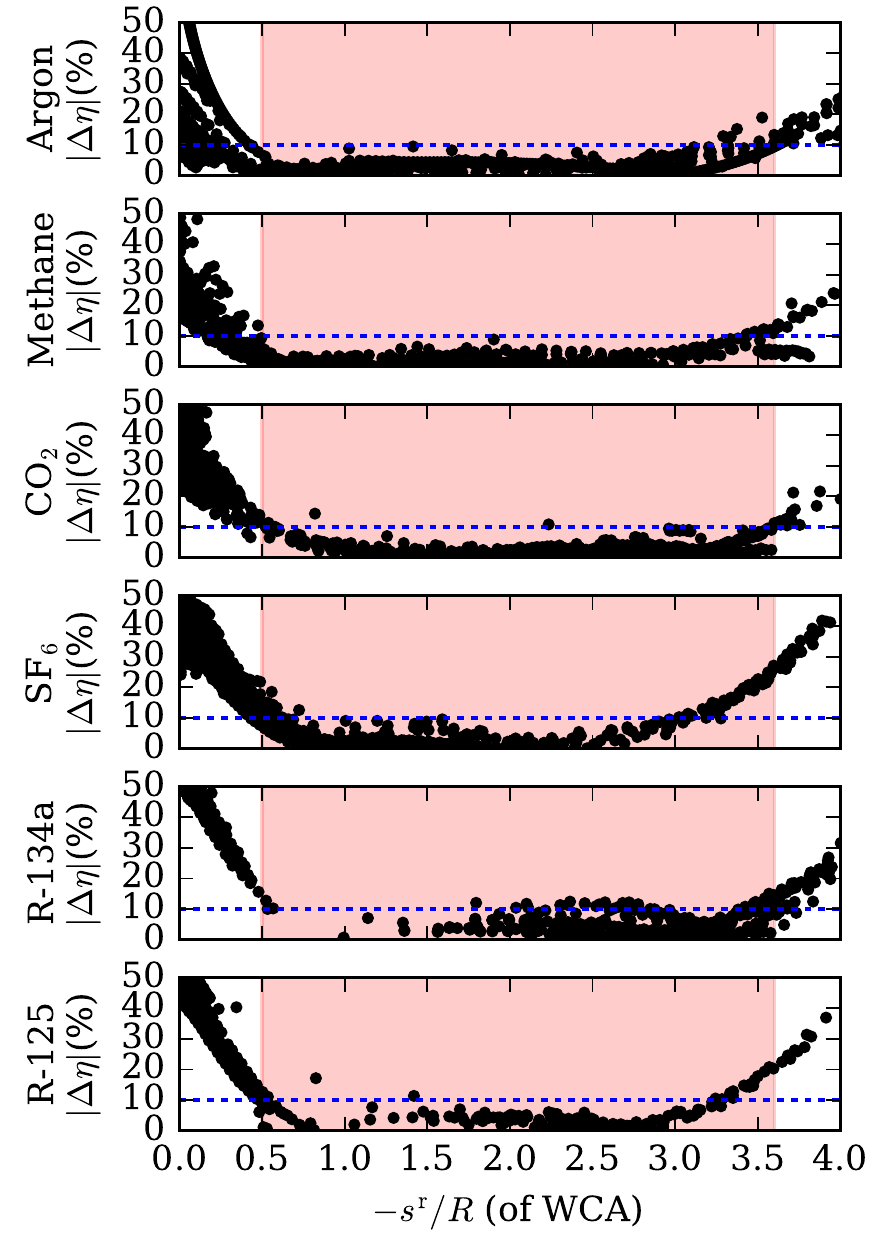}
\caption{Deviations between Rosenfeld-scaled experimental data mapped onto the repulsive WCA potential and experimental data for the non-associating fluids argon, methane, CO$_2$, SF$_6$, R-134a, and R-125.  Absolute deviation is given by $|\Delta\eta|=|(\eta_{\rm fit}/\eta_{\rm exp}-1)\times 100|$.  The colored rectangle is the approximate range of validity of this method, and the dashed line indicates 10\%\label{fig:scaleddeviations_non_hydrogen_WCA}}
\end{figure}

\Cref{fig:scaleddeviations_non_hydrogen_WCA} shows the deviations between the experimental viscosities and the viscosities calculated by the fitted values of $\sigma$ for each of the non-associating fluids described in \cref{fig:simscaled_non_hydrogen_WCA}.  Within the recommended range of validity ($0.5 \lesssim -s^{\rm r}/R \lesssim 3.5$) of the repulsive WCA potential, the deviations are in general less than 10\% within the bulk of the range, except at larger values of $-s^{\rm r}/R$, where the curvature of the repulsive WCA potential results begins to move the correlation away from the experimental data (see \cref{fig:simscaled_non_hydrogen_WCA}).  Within the center of the region of validity, the absolute deviations are in general less than 5\%.

The same exercise was made for all the molecular fluids that a) have a Helmholtz-energy-explicit equation of state available in the NIST REFPROP thermophysical property library \cite{LEMMON-RP10} and b) have experimental liquid viscosity data available in the NIST ThermoData Engine \#103b version 10.1 \cite{TDE-10.1}.  The fluids included in this suite include hydrocarbons, refrigerants, siloxanes, noble gases, fatty-acid methyl esters, etc.  In total, 120 fluids were included in the analysis, with molar masses ranging from 2 kg kmol$^{-1}$ (hydrogen) to 459 kg kmol$^{-1}$ (MD$_4$M).

\begin{figure}[htb]
\centering
\includegraphics[width=3.5in]{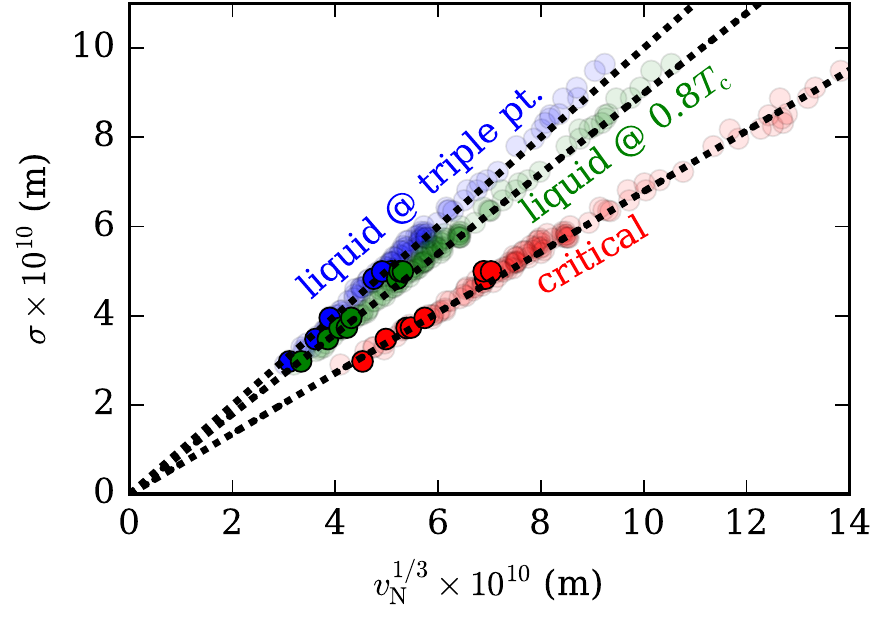}
\caption{Optimized values of $\sigma$ for each fluid for the mapping to the repulsive WCA reference potential for $\varepsilon/\kB=\Tc/1.32$.  The slightly transparent markers correspond to the full set of fluids from NIST REFPROP and with experimental viscosity data from NIST ThermoData Engine \#103b version 10.1, and the solid markers correspond to the fluids selected in \cref{sec:molecularfluids}.  Dashed line for the critical point is given by $\sigma=0.6786(v_{\rm N, crit})^{1/3}$, for the triple point is given by $\sigma=(v_{\rm N, triple})^{1/3}$, and for the saturated liquid at $0.8\Tc$ is given by by $\sigma=0.8984(v_{\rm N, 0.8\Tc})^{1/3}$.    \label{fig:fit_for_WCA_sigma}}
\end{figure}

The fitted values for $\sigma$ are shown in \cref{fig:fit_for_WCA_sigma} as a function of three characteristic volumes, those at the critical point, the liquid at the triple point, and the saturated liquid at $0.8\Tc$. The fact that $\sigma^3$ should be proportional to the critical volume was originally proposed in corresponding states theory \cite{Pitzer-JCP-1939,Liu-CES-1998}, and these data confirm this proposition.  A similar linear relationship between cube root of critical volume and length scale was seen by Liu et al. \cite{Liu-CES-1998} with diffusivity data.  It is remarkable that this behavior holds even for fluids that are associating (ethanol, water, etc.), for which this relationship is not expected to be followed.  In this case the proportionality constant of the critical point volume scaling is approximately 0.7, which is quite different than the value given by the Chung model for extended corresponding states \cite{Chung-IECR-1984,Chung-IECR-1988} of 0.958. An even more remarkable relationship is found when the length-scaling parameter is plotted against the cube root of the volume of the liquid at the triple point; the length scaling parameter is \textit{approximately equal} to  $v^{1/3}_{\rm N, triple}$.  We currently have no theoretical explanation for this behavior.  A third length scale based on the cube root of the volume of the saturated liquid at $0.8\Tc$ also results in a nearly linear functional dependence; this is a more meaningful liquid corresponding states point than the triple point because the latter depends on solid-phase properties.  In the SI (Section \zref{sec:altlengthscales}) additional candidate length scales are further described, including the length scale obtained from Noro-Frenkel universalism \cite{Noro-JPC-2000} and the length scales obtained for $\varepsilon/\kB=\Tc$ and $\varepsilon/\kB=\Tc/0.7$ for each reference potential.

While the mapping between experimental data and model potentials via the residual entropy is fruitful, one challenge is that the repulsive model potentials reach their respective solid-liquid-equilibrium curve at smaller values of $-s^{\rm r}/R$ than the experimental data scaled into simulation units.  The largest value of $\eta^*/[(\rho^*)^{2/3}\sqrt{T^*}]$ available for the repulsive WCA potential is 4.3 for the highest density simulation run of \cite{Heyes-JPCM-2007}. Therefore, the mapped data for $\eta^*/[(\rho^*)^{2/3}\sqrt{T^*}] > 4.3$ represent a metastable extrapolation of the repulsive WCA results into the solid phase that should be considered with suspicion.

\begin{figure}[htb]
\centering
\includegraphics[width=3.3in]{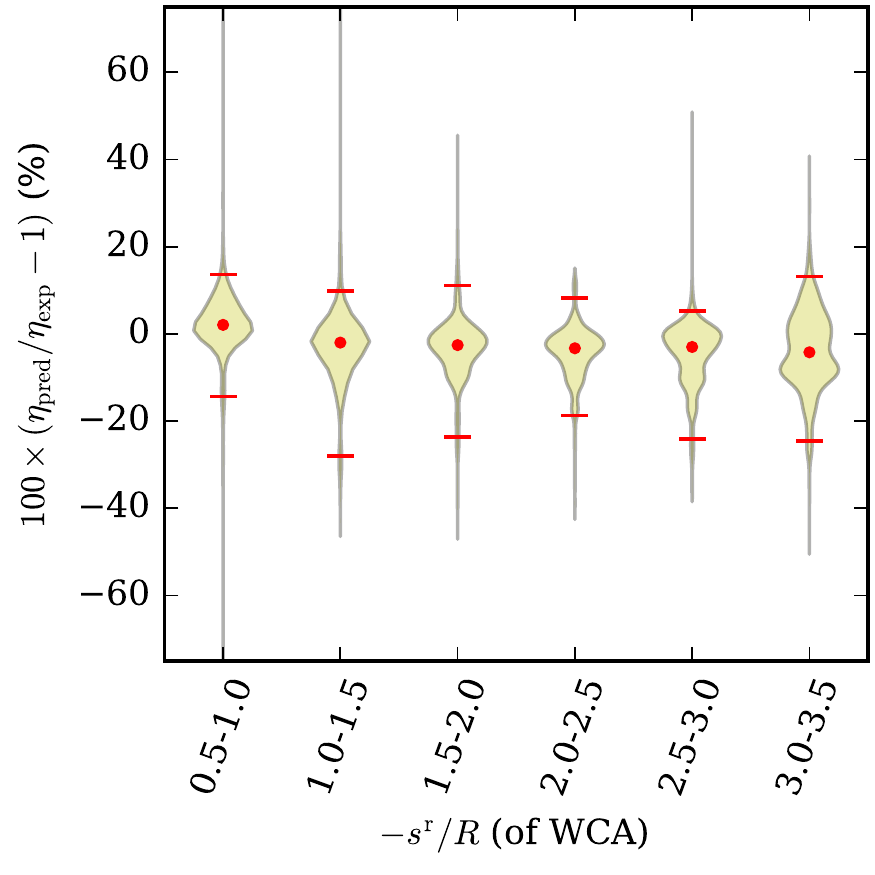}
\caption{Violin plots of the deviations in prediction of viscosity with the optimized values of $\sigma$ for each fluid for the mapping to the repulsive WCA reference potential for non-associating fluids. The range of $-s^{\rm r}/R$ between 0.5 and 3.5 was split into bins of width 0.5.  A violin distribution was constructed (by matplotlib\cite{matplotlib}) for the results in each bin.  The 97.5\% and 2.5\% percentiles are indicated with horizontal lines and the marker is the median value.  Experimental data points for $-s^{\rm r}/R$ greater than 3.5 or less than 0.5 are not shown, and in general correspond to much larger deviations. \label{fig:deviations_TDE_non_hydrogen_fitted_sigma_WCA}}
\end{figure}

Finally, \cref{fig:deviations_TDE_non_hydrogen_fitted_sigma_WCA} presents a set of violin plots for all of the experimental data for the full set of fluids in NIST REFPROP with experimental viscosity data in NIST ThermoData Engine \#103b version 10.1.  Nearly 50,000 experimental data points are included in this collection.  For each fluid, the optimized  value of $\sigma$ for each fluid is used. In the recommended range of validity of the WCA potential ($0.5 \lesssim -s^{\rm r}/R \lesssim 3.5$), 95\% of the data points are predicted within 18.1\%, and the worst median error is 4.2\% for the bin at the largest value of $-s^{\rm r}/R$.  The fully predictive mode, where $\sigma$ is taken from the correlation based upon the volume of the saturated liquid at $0.8\Tc$, as described in \cref{fig:fit_for_WCA_sigma}, results in a poorer representation of the experimental viscosity, as shown in the SI.  In predictive mode, in the same range of validity, 95\% of the data points are predicted within 46.5\%.

\section{Conclusions and Outlook}

This work has demonstrated that the Rosenfeld scaling of viscosity allows the viscosity of pure fluids and model potentials to collapse to nearly monovariate functions of the residual entropy.  This monovariability allows for a mapping from molecular fluid properties onto the properties of the model potentials for non-associating molecular fluids.  Thus, a theoretically grounded approach is demonstrated that connects model potentials and molecular fluids through the residual entropy.  The scaling parameter $\sigma$ is also shown to be nearly proportional to measureable length scales of the molecular fluids.

It is not conclusively shown that the WCA potential is the best possible model potential for the residual entropy corresponding states in viscosity; further study should consider whether other model potentials would be more suitable.  For instance it is seen that the scaled viscosity data in \cref{fig:simscaled_non_hydrogen_WCA} does not have the same curvature as the repulsive WCA potential.  A ``better" potential would more faithfully represent the shape of the viscosity data in these scaled coordinates.

Ultimately, the connection between residual entropy and viscosity stems from the fact that viscosity is primarily governed by the repulsive interactions between molecules.  The structure in the fluid is driven by the repulsive interactions \cite{Weeks-JCP-1971,Barker-JCP-1967,Dyre-JPCM-2016}, so if structure is the determinant of viscosity, and if structure can be quantified by the residual entropy, then it follows that the viscosity should be closely related to the residual entropy.  

There are many molecular fluids for which no experimental viscosity data exist.  This universal scaling approach, along with the scaling parameters of \cref{fig:fit_for_WCA_sigma}, can yield a reasonable estimate for viscosities of heretofore unmeasured fluids, as long as they are not associating.  Or, if a small number of viscosity measurements are available, $\sigma$ could be fit to those data points and the entire liquid viscosity surface accurately predicted within perhaps 20\%.  The mapping of associating fluids onto model potentials remains a challenging endeavor, and worth continued research effort.  

\acknow{\acktext}

\showacknow{} 

\section{Supplemental Material}
\begin{itemize}
\item Detailed information on the data sources for each molecular fluid and model potential
\item Derivations and description of the analysis required for the model potentials
\item Additional figures for completeness
\end{itemize}

\bibliographystyle{myunsrt}
\bibliography{CoolPropBibTeXLibrary,Alexandria,ArnoFluids} 

\end{document}